\documentclass[%
reprint,
 amsmath,amssymb,
 aps,
]{revtex4-2}

\usepackage{graphicx}
\usepackage{dcolumn}
\usepackage{bm}
\usepackage{color}
\usepackage{ulem}

\begin{document}


\title{Random coherent states as a mimic for quantum illumination} 

\author{Thomas Brougham, Nigam Samantaray and John Jeffers}
\affiliation{Department of Physics, University of Strathclyde, John Anderson Building, 107 Rottenrow, Glasgow G4 0NG,
United Kingdom}


\date{\today}

\begin{abstract}
Quantum illumination uses quantum correlations to enhance the detection of an object in the presence of background noise.  This advantage has been shown to exist even if one uses non-optimal direct measurements on the two correlated modes.  Here we present a protocol that mimics the behaviour of quantum illumination, but does not use correlated or entangled modes.  Instead, the protocol uses coherent (or phase-randomized coherent) pulses with randomly chosen intensities.  The intensities are drawn from a distribution such that the average state looks thermal.  Under appropriate conditions, the mimic protocol can perform similarly to quantum illumination schemes that use direct measurements.  This holds even for a reflectance as low as $10^{-7}$. We also present an analytic condition which allows one to determine  the sets of parameters in which each protocol works best.   
\end{abstract}

\maketitle 

\section{Introduction}
\label{sec:intro}

Quantum correlations play a central role in many applications of quantum information \cite{teleport,ekert,ghostimage,noon,sensing}.  For instance, quantum illumination protocols use entangled photons to enhance our ability to detect an object \cite{lloyd2008}.  It has been shown that quantum illumination provides an advantage for object detection in the presence of background noise \cite{lloyd2008,tan2008,barzanjeh2015,jeffers2021}.  These schemes uses the enhanced correlation between two modes to improve the contrast between signal and background light.  Quantum illumination thus allows one to detect objects without resorting to increasing the signal strength.  This can be important in applications where the object might be fragile or when we want our actions to be covert.  When covertness is important, we have an additional constraint: the photon statistics of the signal should be of the same form as the background \cite{covert2020}.  This is not the case for a weak coherent laser source, which has Poissonian statistics \cite{Loudon,mandel}.  However, quantum illumination schemes use states of light where the reduced state of the signal modes can have thermal photon statistics \cite{covert2020,rarity1998,knight,yurke}.

The original quantum illumination protocols required a joint measurement of the photons received from the target and the stored idler modes \cite{lloyd2008,tan2008}.  The performance of the protocol is quantified using the Helstrom bound \cite{helstrom, croke2008} and the quantum Chernoff bound \cite{tan2008,chernoff}.  The requirement to store the photons in the idler mode until the signal returns from the target, increases the experimental difficulty of the protocols. This makes the ideal joint measurement challenging and even if it is known, it may be beyond current experimental methods. Instead, non-optimal joint measurement schemes have been proposed \cite{guha2009}; but these are still experimentally challenging.  For these reasons, alternative quantum illumination protocols have been suggested, which are more experimentally feasible \cite{Lopaeva2013,England2019,yang2021,yangspie,yang2022}.  In these schemes, the idler mode is measured separately from the signal mode.  This removes the requirement for storage or for joint measurements.  Furthermore, rather than using an optimal measurement, a simpler measurement is made using threshold detectors.  The correlation between the idler and signal modes means that if an object is present, then there should be correlations between detection of the idler photons and those reflected from the object onto the signal detector.  In contrast, if no object is present, then the there will be no correlation between the idler and signal detectors.  Despite the simplicity of these schemes, they have been shown to outperform a weak coherent source \cite{England2019,yang2022}.  Furthermore, these schemes can still satisfy the covertness condition if they use two-mode squeezed vacuum (TMSV) states \cite{covert2020,yang2022}.

For real world applications, the expected fraction of photons reflected from an object will be very low \cite{radar,lidar}.  This would necessitate working with sources that generate a slightly higher mean photon flux.  This can be accomplished more straightforwardly with a source that produces either coherent states or phase-randomized coherent states.  Furthermore, the requirement to use either an entangled source \cite{tan2008} or correlated photon source \cite{Lopaeva2013,England2019} could increase the complexity and cost of any commercial device.  Both of these reasons point towards the practical advantages of using a coherent source.  However, as stated, these sources does not perform as well as protocols that use a two-mode squeezed vacuum state.  Furthermore, the photon statistics from a single-mode coherent state are Poissonian and thus can be easily differentiated from the thermal background by performing a measurement of the 2nd order temporal coherence, $g^{(2)}(\tau)$ \cite{Loudon,mandel}.

Here we propose a scheme that uses coherent states with random intensities to mimic the behaviour of quantum illumination schemes.  We show that the performance of this scheme is better than one that uses fixed coherent pulses; and under certain conditions is as good as existing illumination protocols. Furthermore, if the intensities of the coherent states are chosen with the correct probability distribution, then the protocol will satisfy the covertness condition by producing average photon statistics which are thermal.  

The outline of the paper is as follows.  In section \ref{sec:qi} we describe a specific quantum illumination scheme in detail.  Understanding this scheme will help to motivate the mimic protocol.  The random coherent state mimic protocol is outlined in section \ref{sec:mimic}.  We present results in section \ref{sec:res} and compare the relative performance of each protocol.  This is achieved by using a Bayesian estimation procedure and performing Monte Carlo simulations of the setup \cite{inforef}. In section \ref{sec:compare} we present an analytic criterion to determine when the mimic protocol will perform better, on average, than an existing quantum illumination protocol.  Finally, we discuss the results in the conclusions.

\section{Description of quantum illumination}
\label{sec:qi}
The motivation for the mimic protocol comes from comparison with an existing quantum illumination protocol, outlined in \cite{yang2022}.  In this scheme the idler mode is not stored, but is instead measured.  Furthermore, threshold detectors are used to measure the signal mode.  As such, we will refer to this scheme as a {\it direct measurement} protocol.  In this section we recap this scheme and explain the key insights that we exploit in the mimic protocol.  This section will also serve as an opportunity to describe the physical model used in both schemes. 

For the direct measurement scheme outlined in \cite{yang2022}, we use a TMSV state \cite{mmqo}
\begin{equation}
\label{tmsv}
|\Psi\rangle_{I,S} = \frac{1}{\sqrt{1+\bar{n}}} \sum_{n=0}^{\infty} \left(\frac{\bar{n}}{1+\bar{n}} \right)^{n/2}|n\rangle_I |n\rangle_S,
\end{equation}
where $\bar{n}$ is the mean photon number in each mode, $|n\rangle$ is an $n$-photon Fock state and the subscripts $I$ and $S$ respectively denote idler and signal modes.  This state can be generated experimentally using nondegenerate spontaneous parametric down-conversion \cite{exp1,exp2,exp3}.  The signal mode is transmitted to where we believe a reflecting target object might be, while the idler mode is measured.  If there is an object, then some photons will be scattered onto the mode of the signal detector.  Both the idler and signal detectors are threshold detectors with efficiency $\eta$.  The idler detector is assumed to be shielded from background photons and could, if required, be gated on and off such that the dark count rate is so small that we can neglect it.  In contrast, the signal detector will receive background photons.  The state of these photons is given by a thermal state \cite{mmqo}
\begin{equation}
\label{thermal1}
\hat\sigma_{\bar{m}}=\frac{1}{1+\bar{m}} \sum_{n=0}^{\infty} \left(\frac{\bar{m}}{1+\bar{m}} \right)^n |n\rangle \langle n|,
\end{equation}
where $\bar{m}$ is the mean number of photons in the thermal state.  Notice that if we trace over the idler mode of the TMSV state, then the reduced state of the signal mode is a thermal state with mean photon number $\bar{n}$.  The photon statistics for the signal mode are of the same form as the thermal background.  In particular, if someone performed a measurement of the 2nd order temporal coherence, $g^{(2)}(\tau)$, for $\tau=0$, they would obtain the same value as for the background \cite{Loudon,mandel}.  The TMSV state will thus satisfy the covertness condition, provided $\bar{n}$ is not too large relative to the background.  

\begin{figure}
\center{\includegraphics[height=5cm]
{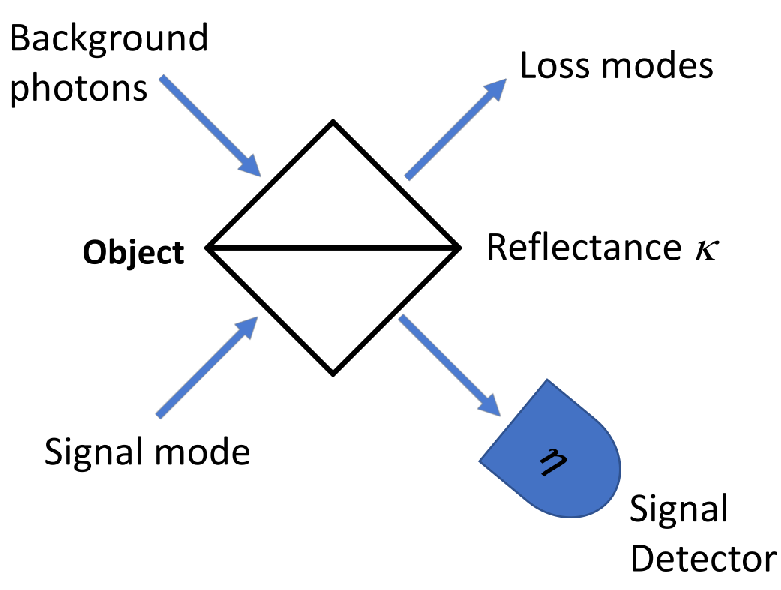}}
\caption{A figure showing the model for photons scattering off an object.  We describe the object by a beam-splitter with reflectance $\kappa$.  Photons that are not reflected into the mode of the detectors are grouped together to form a loss mode, which is traced out.  The effects of background photons is described by mixing a thermal state with the signal mode.}
\label{fig:model}
\end{figure}

If an object is present, then some signal photons will scatter towards the signal detector. The majority of photons, however, will scatter into modes which are not detected.  Mathematically, we can group these modes into a single loss mode, which we trace out. Let $\kappa$ be the probability that a single signal photon is scattered into the mode of the signal detector.  We model the object as a beam-splitter with reflectance $\kappa$, as illustrated in figure \ref{fig:model}.  In the other input mode, we inject a thermal state with mean photon number $\bar{n}_{B}/(1-\kappa)$.  If there is no object, then none of the signal photons will scatter onto the signal detector.  The only photons incident on the detector are from the thermal background.  The mean number of photons per time-bin incident on the signal detector is $\bar{n}_B$.  We thus refer to $\bar{n}_B$ as the mean number of background photons per time-bin.  The effect of dark counts in the signal detector can be included within $\bar{n}_B$ as explained in the appendix B.  

The detection of the idler mode is made with an inefficient threshold detector.  Due to the strong photon number correlations between the signal and idler beams inherent in equation (\ref{tmsv}), detection of light in the idler mode conditions the signal mode to have a mean photon number greater than $\bar{n}$. In contrast, failure to detect light in the idler mode conditions the signal mode to have a mean photon number which is less than $\bar{n}$.  On average, the signal mode is in a thermal state with a mean photon number $\bar{n}$.  However, provided we record the results of the idler measurement, then we know the conditional state, which gives additional information to use in the estimation process.  The procedure can be improved further by using a balanced optical network to split the idler mode evenly onto $N$ identical threshold detectors \cite{yang2022}.  In this case there is a nonzero probability for more than one detector to fire that, when it occurs, gives a greater enhancement for the mean photon number in the signal mode. Nevertheless, the average state of the signal mode is still a thermal state with mean photon number $\bar{n}$. For additional information on the photon statistics of the signal mode, see appendix D.

The key idea for this protocol is that conditioning on the idler mode changes the state of the signal mode.  Our knowledge of the outcome for the idler measurement improves our ability to determine whether an object is present or not. However, without knowledge of the idler measurement, the no-signaling theorem \cite{nosignal} implies that the signal mode will, on average, be a thermal state.

\section{Mimic lidar with coherent states}
\label{sec:mimic}
In this section we outline a mimic lidar scheme that uses random coherent states. If we were to send a fixed, weak coherent state, we would not satisfy the covertness condition, as the photon statistics would be Poissonian and thus could be detected by an appropriate measurement. Furthermore, the direct measurement scheme outperforms a fixed coherent state when $\bar{n}_B$ is large relative to the mean number of signal photons $\bar{n}$, which for a coherent state $|\alpha\rangle$ is $\bar{n}=|\alpha|^2$ \cite{yang2022}.

Rather than transmitting a fixed coherent state, we instead randomly pick coherent states with different complex amplitudes, $\alpha$. To satisfy the covertness condition, we need the average ensemble to be a thermal state.  The $P$-representation of a thermal state with mean photon number $\bar{m}$ is \cite{mandel}
\begin{equation}
\label{thermal2}
\hat\sigma_{\bar{m}}=\frac{1}{\pi\bar{m}}\int e^{-|\alpha|^2/\bar{m}} |\alpha\rangle\langle\alpha| d^2\alpha.
\end{equation}
This density operator can be prepared by randomly generating a coherent state $|\alpha\rangle$ with probability density (over a complex space) $\exp(-|\alpha|^2/\bar{m})/(\pi\bar{m})$.  An experimental procedure for this has been demonstrated in a protocol for covertly distributing information \cite{ghostdisplace}. This suggests the following protocol.

In each time-slot, we randomly pick an amplitude, $\alpha=|\alpha|e^{i\theta}$, for a coherent state, where we use the probability distribution $p_{\bar{n}}(\alpha)=\exp(-|\alpha|^2/\bar{n})/(\pi\bar{n})$.  This is equivalent to picking the phase, $\theta$, uniformly and then picking a mean photon number, $|\alpha|^2$ from the distribution $p_{\bar{n}}(|\alpha|^2)=\exp(-|\alpha|^2/\bar{n})/\bar{n}$.  The mean photon number, $|\alpha|^2$, is recorded for each time-slot.  We do not need to record the phase of $\alpha$ as this will not feature in the conditional probabilities for the detector to fire. In any case the absolute phase may not be physically meaningful if we use a laser to provide our state as theoretical considerations suggest that ascribing such a property to a laser state may be a ``convenient fiction'' \cite{molmer, peggjeff}. We generate a series of single-mode coherent pulses with the chosen amplitudes and record whether the signal detector fires in each time-slot.  We know the list of mean photon numbers, $|\alpha|^2$, and together with the measurement record it is used to estimate the probability that the object is present.  This is achieved using the following Bayesian approach.  

For the separate cases where the object is present or not, we calculate the probability for the detector to fire or not, given the pulse had intensity $|\alpha|^2$.  When there is no object, the probability to not see a click is
\begin{equation}
\label{noobject}
P(0|\bar{O},|\alpha|^2)=\text{Tr}[\hat\sigma_{\bar{n}_B}\hat\Pi_0]=\frac{1}{1+\eta\bar{n}_B},
\end{equation}
where $\bar{O}$ denotes no object and $\hat\Pi_0=\sum_n (1-\eta)^n|n\rangle\langle n|$ is the positive operator-valued measure for the no-click result. The probability for the detector to fire is $P(1|\bar{O},|\alpha|^2)=1-P(0|\bar{O},|\alpha|^2)$.  When an object is present, the probability to not register a click is
\begin{equation}
\label{eq:detect}
P(0|O,|\alpha|^2)=\frac{1}{1+\eta\bar{n}_{B}}\exp\left(\frac{-\eta\kappa|\alpha|^2}{1+\eta\bar{n}_{B}}\right),
\end{equation}
where $O$ denotes that an object is present.  See appendix A for a derivation.  The probability to register a click is $P(1|O,|\alpha|^2)=1-P(0|O,|\alpha|^2)$. We do not know whether the object is actually present; we only have our set of consecutive experiments to guide us. So we base our decision on the Bayesian posterior probability that an object is present, which we can calculate based on the outcomes of our measurements. The reason for adopting a Bayesian approach is that it easily allows one to incorporate any prior information about whether an object is likely to be present at a given location.

Let ${\bf x}^{(r)}$ denote an array of the first $r$ measurement outcomes for the detector and let ${\bf n}^{(r)}$ be an array of the first $r$ mean photon numbers used when preparing the pulses.  It will be convenient to introduce some notation, let $\Omega_r=\{{\bf x}^{(r)},{\bf n}^{(r)}\}$.  After each pulse, we update the posterior probability for the target object to be present using Bayes' rule.  After the $r$-th measurement, the probability for the object to be present is
\begin{eqnarray}
\label{bayes}
&&P(O|{\bf x}^{(r)},{\bf n}^{(r)})=P(O|\Omega_r)=\\
&&\frac{P(x_r|O,|\alpha|_r^2)P(O|\Omega_{r-1})}{P(x_r|O,|\alpha|_r^2)P(O|\Omega_{r-1})+P(x_r|\bar{O},|\alpha|_r^2)P(\bar{O}|\Omega_{r-1})}, \nonumber
\end{eqnarray}
where $x_r$ is the outcome of the $r$-th measurement and $|\alpha|_r^2$ is the mean photon number of the $r$-th pulse.  Initially, we have no knowledge of whether an object is present. We account for this by using equal prior probabilities, i.e., $P(O)=P(\bar{O})=1/2$. We might, of course, make a different prior choice if we have a greater belief that no object is present initially.

After $N$ measurements $P(O|\Omega_N)$ is the posterior probability for the object to be present.  This will depend on the set of measurement outcomes ${\bf x}^{(N)}$ and the set of intensities chosen ${\bf n}^{(N)}$.  The effectiveness of the protocol can be evaluated by performing a Monte Carlo simulation to evaluate the average performance \cite{inforef}.  This entails performing a random simulation of the experiment.  For this simulation we describe the system by the model shown in figure (\ref{fig:model}).  We randomly generate a set of mean photon numbers for each pulse.  These are used with Eq. (\ref{eq:detect}) to obtain a set of simulated measurement outcomes consistent with those produced if a target were present.  These outcomes are fed into (\ref{bayes}) to find the probability for the object to be present after each pulses.  This gives $P(O|\Omega_N)$ for one possible set of measurement outcomes. We would then run this many times and average $P(O|\Omega_N)$. 

We have explained the mimic protocol for coherent states.  However, the phase of the coherent states is not important.  One could thus replace coherent states with phase-randomized coherent states. If we set the mean photon number to $\lambda$, then phase randomized coherent states have the form 
\begin{equation}
\hat\rho_{\lambda}=\frac{1}{2\pi}\int^{2\pi+\theta_0}_{\theta_0}|\sqrt{\lambda}e^{i\theta}\rangle\langle\sqrt{\lambda}e^{i\theta}| d\theta.
\end{equation}
By writing (\ref{thermal2}) in polar coordinates and performing the phase integration, we can verify that 
\begin{equation}
\hat\sigma_{\bar{m}}=\frac{1}{\bar{m}}\int^{\infty}_0 e^{-\lambda/\bar{m}}\hat\rho_{\lambda}d\lambda.
\end{equation}
The protocol is then exactly the same as before.  In particular, the probabilities (\ref{noobject}) and (\ref{eq:detect}) are the same, but with $\lambda=|\alpha|^2$.

\section{Results}
\label{sec:res}
To evaluate the average performance of the mimic protocol, we compare it to other approaches.  We look at the quantum illumination protocol discussed in section \ref{sec:qi} and also compare it to coherent pulses with fixed amplitudes.  The latter protocol does not satisfy the covertness condition as it has Poissonian photon statistics.  Nevertheless, transmitting fixed coherent states is straightforward and it is thus sensible to compare the performance of more complicated schemes against this simple approach.

For a fair comparison between the three approaches, we use the same mean photon number for the signal mode of each protocol.  Recall that for the direct measurement and mimic protocols, the mean photon number will vary pulse to pulse.  For these protocols $\bar{n}$ denotes the averaged mean photon number in the signal mode.  For the fixed coherent state $|\alpha\rangle$, then $\bar{n}=|\alpha|^2$.  For all protocols, we perform a Monte Carlo simulation to obtain the average performance of each approach.  
For more details on performing a Monte Carlo simulation for the direct measurement protocol, see \cite{yang2022}. The simulation for the fixed coherent state is the same as outlined in the last section for the mimic scheme, but where now the amplitude of the coherent state is fixed.

In practice, we can use a source with high repetition rate of 100MHz or 1GHz. In the latter case, we transmit a million pulses to the target in a millisecond.  A large number of pulses can thus be used while still having a low acquisition time, which will allow detection of objects of low reflectivity.  However, performing averaging over many realisations of simulations that each cover large numbers of pulses is time consuming.  To reduce the number of pulses we investigate the regimes where $\kappa=0.1$.  This corresponds to an object which is expected to be relatively close to the source.  The reduced number of pulses needed for larger values of $\kappa$ allows for a investigation of results with very high levels of background counts, without the need for advanced computational resources.  This is of particular importance as quantum illumination was initially conceived to solve the problem of object detection in environments with high background, while using relatively low signal powers.  

\begin{figure}
\center{\includegraphics[width=8cm]
{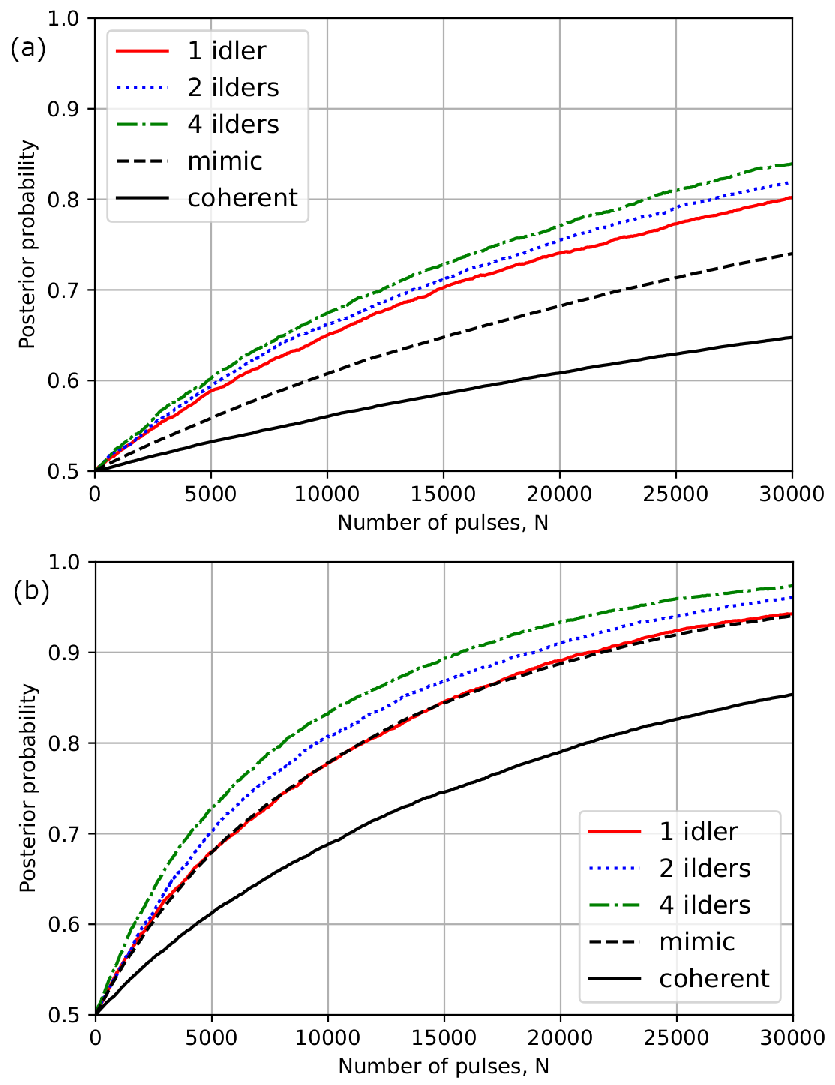}}
\caption{A plot of the probability for the object to be present given we have transmitted $N$ signals.  All plots are for the case where an object is present, $\eta=0.9$, $\bar{n}_B=3.0$ and $k=0.1$. Figure (a) is for a mean photon number of 0.5 photons per pulse, while (b) is for 1.0 photons per pulse. In both figures, the dotted and dashed line(green online), dotted (blue online) and solid gray (red online) curves refer to the direct measurement scheme where: solid (red) is for a single idler detector, dotted (blue) is for 2 idler detectors and dotted and dashed (green) is for 4 idler detectors.} The dashed black curve is for the mimic protocol and the solid black curve is a fixed coherent state.  The black and dashed curves have been averaged over 8000 runs, while the other curves have been averaged over 4000 runs.
\label{fig:plot1}
\end{figure}

In figure \ref{fig:plot1} we plot the probability for an object to be present, given we have used $N$ pulses.  For all curves, we have performed Monte Carlo simulation for $\bar{n}_B=3.0$, $\eta=0.9$ and where an object is present with $\kappa=0.1$.  For the direct measurement scheme, we show results for a single idler detector (solid gray, red online), two idler detectors (dotted, blue online) and four idler detectors (dotted and dashed, green online).  We compare this with the mimic protocol (black dashed) and a fixed coherent state (solid black).  In figure (a), the mean photon number of each pulse is $\bar{n}=0.5$, which is six times lower than the number of background photons $\bar{n}_B$.  We see that the mimic system does not perform as well as the direct measurement scheme, but does significantly better than the scheme with a fixed coherent state.  In figure (b), the mean photon number for each pulse is $\bar{n}=1.0$.  We now see that the mimic protocol almost performs as well as the direct measurement protocol with a single idler detector.  Notice that in this case, $\bar{n}_B$ is still three times larger than $\bar{n}$.  Both plots (a) and (b) demonstrate that mimic protocol can outperform a coherent state of fixed amplitude.  In plot (b) after 30000 shots the mimic and the single idler detector outperform the coherent state protocol with an increase in posterior probability of almost
0.1. A better way of quantifying this out-performance, however, is to use the number of shots required to reach a particular level of confidence. On the right hand edge of (b) the coherent state protocol reaches a probability of around 0.85 in 30000 shots. The mimic and 1-photon idler schemes reach this confidence level in half the time.

As $\bar{n}$ increases we see that all protocol improve.  However, in relative terms, there is a greater increase in the performance for the mimic protocol.  One reason for this is that in the mimic protocol we will sometimes transmit a coherent state with a relatively large coherent amplitude.  When $\bar{n}$ increases, there is a higher probability to pick a coherent state with a large amplitude.  This can be illustrated with the following simple example.  The probability to send a pulse with $|\alpha|^2\ge 2$ is 0.018 for $\bar{n}=0.5$, which increases to 0.135 for $\bar{n}=1.0$.  These large amplitude pulses give us more information .  This is confirmed by looking at the individual (unaveraged) trajectories.  We observe that detection of pulses with large intensities tend to yield a greater increase in probability that the object is present.  For more information on this effect and the photon statistics of the ensemble, see appendix D. Numerical investigations for $\bar{n}=2.0$ and the same values of $\eta$, $\bar{n}_B$ and $\kappa$ as was used in figure \ref{fig:plot1}, found that the mimic protocol performed almost as well the 4 idler detector, direct measurement scheme.  This confirms that in relative terms, the mimic protocol performs better at higher mean photon numbers, than the direct measurement protocol.

\begin{figure}
\center{\includegraphics[width=8cm]
{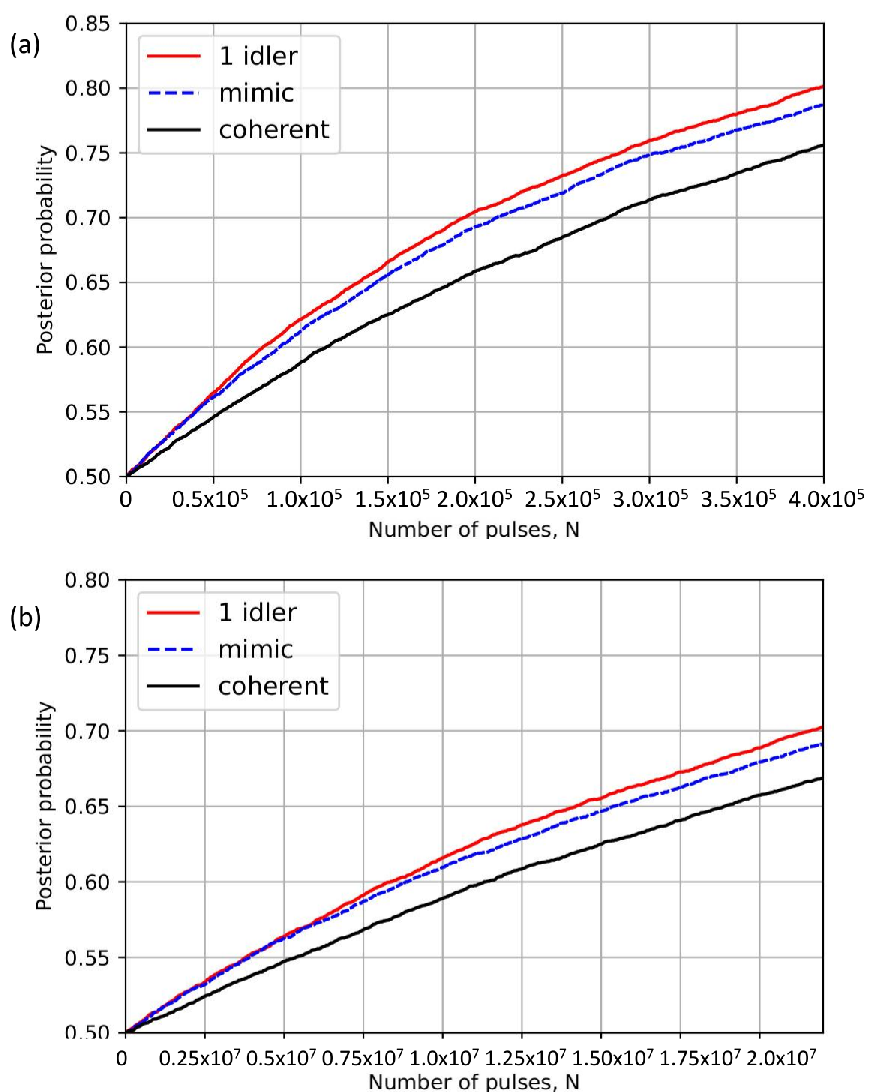}}
\caption{A plot of the probability for the object to be present given we have transmitted $N$ signals.  All plots are for the case where an object is present, $\eta=0.9$ and $\bar{n}=1.0$. Figure (a) is for $\kappa=10^{-5}$ and $\bar{n}_B=5.56\times 10^{-6}$, while (b) is for $\kappa=10^{-7}$ and $\bar{n}_B=5.56\times 10^{-8}$. In both figures, the gray (red online) curve refers to the direct measurement scheme with a single idler detector, the dashed (blue online) curve is for the mimic protocol and the solid black curve is a fixed coherent state.  All curves have been averaged over 8000 runs.}
\label{fig:plot2}
\end{figure}

In many practical situations, the target will not be close to the detection system. The effective reflectivity of the object will be proportional to the solid angle subtended by the detector at the target.  If we include this effect in $\kappa$, it will be low and thus a very small fraction of signal photons will reach the detector.  We now consider the performance of the protocol in the limit of low reflectivity. In figure \ref{fig:plot2} we plot the averaged probability for the object to be present given we have results for $N$ pulses.  For all curves we have performed Monte Carlo simulations for $\eta=0.9$ and a mean photon number of $\bar{n}=1.0$ for each pulse.  Plot (a) is for $\kappa=10^{-5}$  and $\bar{n}_B=5.56\times 10^{-6}$, while (b) is for $\kappa=10^{-7}$ and $\bar{n}_B=5.56\times 10^{-8}$. We show results for the direct measurement protocol with a single idler detector (gray, red online), the mimic protocol (dashed) and the fixed coherent state (black).  All curves are averaged over 8000 random runs. 

To gain some physical perspective on values of $\bar{n}_B$ used in figure \ref{fig:plot2}, we compare them to detector dark counts.  Equation (\ref{modified_background}) of the appendix B allows us to incorporate the dark count probability in $\bar{n}_B$.  Suppose the system is operating in an environment where we can ignore stray light, then from Eq. (\ref{modified_background}) we can set $\bar{n}_B=P_D/\eta(1-P_D)$, where $P_D$ is the dark count probability.  A value of $\bar{n}_B=5.56\times10^{-8}$, corresponds to $P_D=5.00\times10^{-8}$.  For a laser source with a repetition rate of 1 GHz, we would expect approximately 50 dark counts per second for this value of $P_D$.  Similarly, $\bar{n}_B=5.56\times10^{-6}$, with a source repetition rate of 1 GHz, corresponds to slightly over 5000 counts per second.

Figure \ref{fig:plot2} again shows that the mimic protocol performs better than a fixed coherent state.  Furthermore, the mimic protocol is slightly worse, but still close to the performance of the direct measurement scheme with a single idler detector.  Figure \ref{fig:plot2} (b) shows that even for a reflectance of $\kappa=10^{-7}$, {\it both} the mimic and direct measurement protocols give an advantage over fixed coherent states.  Quantum illumination schemes with direct measurements still provide an advantage for low reflectivities. Note also, that figure \ref{fig:plot2} (b) shows that less than $3\times 10^{7}$ pulses are needed for reasonable confidence that an object is present.  For a pulsed laser source with a 1-GHz repetition rate, this corresponds to an acquisition time of less than 0.03s.  

\begin{figure}
\center{\includegraphics[width=8cm]
{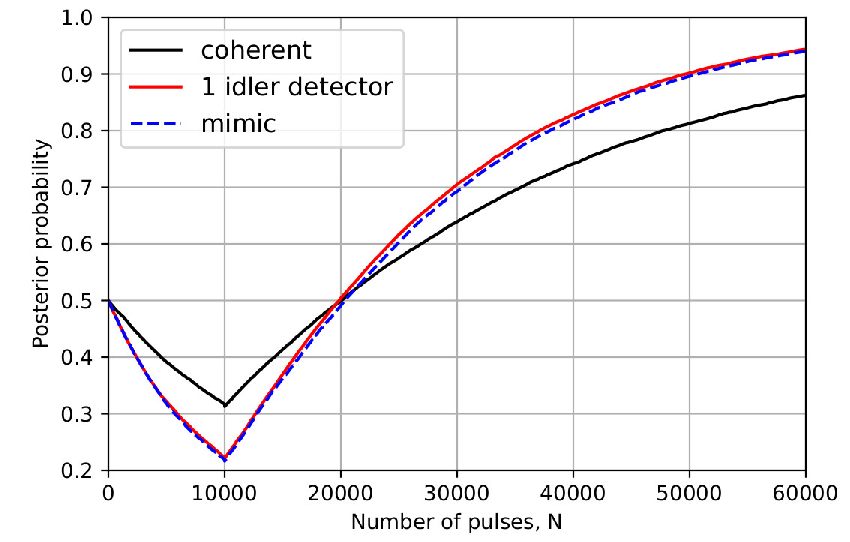}}
\caption{A plot of the probability for the object to be present given we have transmitted $N$ signals.  All plots are for $\eta=0.9$, $\bar{n}_B=3.0$ and $\kappa=0$ for $N<10000$, and $\kappa=0.1$ for $N\ge 10000$. The gray (red online) curve refer the direct measurement scheme with a single idler detector, the dashed (blue online) curve is for the mimic protocol and the solid black curve is a fixed coherent state.  All curves have been averaged over 8000 runs.}
\label{fig:plot_td}
\end{figure}

Thus far we have considered situations where an object is always present.  However, the approach that we have taken applies also to more general situations, such as an object not being present initially but appearing during the process of illuminating an area.  To simplify the analysis, we assume the object appears within the time between pulses and is visible after the 10,000$^{\mbox{th}}$ pulse. The flexibility of the Bayesian approach means that we do not need to change our estimation procedure for this situation.  The results of a Monte Carlo simulation are plotted in figure \ref{fig:plot_td} for the parameters $\bar{n}_B=3.0$, $\eta=0.9$ and $\kappa=0.1$ when the object is present and zero before that. The solid black line corresponds to a fixed coherent state, the dashed (blue online) curve is for the mimic protocol, while the solid gray (red online) curve is for the direct measurement scheme with a single idler detector.  The performance of the mimic protocol is very close to the direct measurement scheme, while both are better than using a fixed coherent state.  In particular, we see that the mimic scheme performs better than a fixed coherent state at registering the absence of the object and then responds quicker to the object's appearance.  For instance, after 10,000 pulses, the mimic scheme has an averaged probability of $\approx 0.22$, while for the fixed coherent state, the probability is $\approx 0.32$.

\begin{figure}
\center{\includegraphics[width=8cm]
{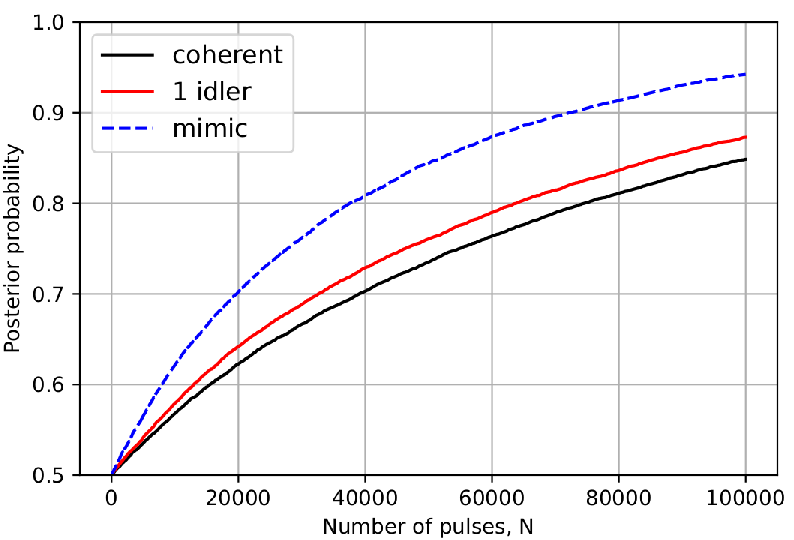}}
\caption{A plot of the probability for the object to be present given we have transmitted $N$ signals.  All plots are for the case where an object is present, $\eta=0.2$, $\kappa=0.02$ and $\bar{n}=\bar{n}_B=20.0$. The dashed curve (blue online) is for the mimic protocol, the solid black curve is a fixed coherent state and the gray (red online) curve is for the threshold protocol with a single idler detector.  All curves have been averaged over 8000 runs.}
\label{fig:classical}
\end{figure}

An important feature of the mimic protocol is that it uses coherent pulses and as such, it is straightforward to increase the intensity.  It can thus be used in high intensity regimes that would be unsuitable for direct measurement schemes.  This can be important in many applications where we have higher backgrounds or high losses.  We illustrate this with an example where $\kappa=0.02$, $\eta=0.2$, $\bar{n}_B=20.0$ and an object is present.  In figure \ref{fig:classical} we plot the posterior probability to detect an object for $\bar{n}=20.0$. The figure shows that the mimic protocol provides an advantage over both a fixed coherent state and the direct measurement scheme with a single idler detector.  This is important as it illustrates that the mimic protocol has applications outwith the quantum regime of low photon number.

Figure \ref{fig:classical} illustrates that the mimic protocol can significantly outperform direct measurement schemes when the mean photon number becomes relatively large.  For instance, to achieve a posterior probability of 0.8 requires $N\approx 37,000$ pulses for the mimic protocol, while the 1 idler detector direct measurement scheme requires $N\approx 63,000$ pulses.  Another example of this is, the example mentioned earlier, with $\kappa=0.1$, $\eta=0.9$, $\bar{n}_B=3.0$ and $\bar{n}=2.0$, where we found that the mimic protocol's relative performance was almost as good the direct measurement protocol with 4 idler detectors. These examples show that there exists regimes where the mimic protocol can outperform the direct measurement protocol.

\section{Comparison of protocols}
\label{sec:compare}
An obvious question to ask is: for a given set of system parameters, when should we use each protocol?  To fully answer this question would require one to consider the cost and complexity of each protocol.  Instead, we will here focus on the simpler task of determining when each protocol performs best.  We achieve this by finding an analytic criterion for when the mimic protocol is better than the direct measurement protocol, using a single idler detector.   The approach is to look at the averaged posterior probability for a single measurement.  Using Eq. (\ref{bayes}) we find the probability for an object to be present after a single measurement outcome.  This is averaged over all possible measurement outcomes and choices for the pulse intensity.  The same is done for the direct measurement protocol, where now we average over both the idler and signal measurements.  The averaged probabilities are then compared to asses the performance of each protocol.  In all the following calculations, we assume that an object is present and take the initial prior probabilities to be $P(O)=P(\bar{O})=1/2$.

For notational simplicity, we set $|\alpha|^2=\lambda$.  The probability for the object to be present after we obtain the measurement outcome $x$ is $P(O|x,\lambda)$, which can be calculated using (\ref{bayes}).  The average of this over all possible measurement outcomes and pulse choices, $\mathcal{P}^{RC}_{\bar{n}}(O)$, is 
\begin{equation}
\label{prc_def}
\mathcal{P}^{RC}_{\bar{n}}(O)=\int^{\infty}_0 \sum_{x=0}^1 p_{\bar{n}}(\lambda)P(O|x,\lambda)P(x|O,\lambda)d\lambda,
\end{equation}
where $RC$ stands for random coherent and $p_{\bar{n}}(\lambda)=\exp(-\lambda/\bar{n})/(\bar{n})$ is the probability to pick a pulse with mean photon number $\lambda$.  The evaluation of the integral for the two terms in the sum can be performed using the integral representations of the standard hypergeometric function and the Harmonic function \cite{AA}.  For details, see appendix C.  We find that 
\begin{eqnarray}
\label{rc_average}
&&\mathcal{P}^{RC}_{\bar{n}}(O)=\frac{1}{2\eta \kappa\bar{n}}\left[H\left(\frac{1}{2}+\frac{\beta}{2}\right)-H\left(\frac{\beta}{2}\right)\right]\nonumber\\
&+&\beta C\Big[\frac{\,_2 F_1(1,\beta;\beta+1;A)}{\beta}-2\frac{\,_2 F_1(1,\beta+1;\beta+2;A)}{(\beta+1)(1+\eta\bar{n}_B)}\nonumber\\
&+&\frac{\,_2 F_1(1,\beta+2;\beta+3;A)}{(\beta+2)(1+\eta\bar{n}_B)^2} \Big],
\end{eqnarray}
where $H(x)$ is the Harmonic function, which is equal to Euler's constant plus the derivative of the natural log of the gamma function \cite{AA}.  The constants $A$, $\beta$ and $C$ are:
\begin{eqnarray}
A&=&\frac{1}{1+2\eta\bar{n}_B}\nonumber\\
\beta&=&\frac{1+\eta\bar{n}_B}{\eta\kappa\bar{n}}\nonumber\\
C&=&\frac{1+\eta\bar{n}_B}{1+2\eta\bar{n}_B}.
\end{eqnarray} 
For the direct measurement protocol, we average over all possible outcomes for the idler and signal detectors.  Let $s$ and $i$ denote the outcomes for the signal and idler detectors respectively.  The average probability after a single set of measurements is  
\begin{eqnarray}
\label{ent_average}
\mathcal{P}^{DM}_{\bar{n}}(O) &=& \sum_{i,s} P(i,s|O) P(O|i,s)\nonumber\\
&=&\sum_{i,s}p_{\bar{n}}(i)\frac{P(s|i,O)^2}{P(s|i,O)+P(s|i,\bar{O})}, 
\end{eqnarray}
where $p_{\bar{n}}(i=0)=(1+\eta\bar{n})^{-1}$ and $p_{\bar{n}}(i=1)=1-p_{\bar{n}}(i=0)$ are the probabilities of obtaining the outcome $i$ for a measurement on the idler mode.  The detection probability for no object present, $P(s|i,\bar{O})$, is the same as (\ref{noobject}).  The conditional probabilities $P(s|i,O)$ are calculated in \cite{yang2022}.  For no detection, i.e  $s=0$, the probabilities are
\begin{eqnarray}
P(0|0,O)=\frac{1+\eta\bar{n}}{1+\eta(\bar{n}+\bar{n}_B)+\eta\bar{n}\left[\eta\bar{n}_B+(1-\eta)\kappa\right]},\nonumber\\
P(0|1,O)=\frac{1}{\eta\bar{n}}\left(\frac{1+\eta\bar{n}}{1+\eta(\bar{n}_B +\kappa\bar{n})}-\frac{1}{1+\eta(\bar{n}_B +\kappa\bar{m}_{10})}\right),\nonumber\\
\end{eqnarray}
where $\bar{m}_{10}=\bar{n}(1-\eta)/(1+\eta\bar{n})$.

As an example of how one can use Eqs (\ref{rc_average}) and (\ref{ent_average}), with the parameters $\bar{n}_B=3.0$, $\eta=0.9$ and $\kappa=0.1$ (i.e. figure \ref{fig:plot1} (b)).  For these values, we find that $\mathcal{P}^{RC}_{\bar{n}}(O)>\mathcal{P}^{DM}_{\bar{n}}(O)$ when $\bar{n}\ge 1.04$.  If we keep $\bar{n}_B$ and $\eta$ the same, but change the reflectance to $\kappa=10^{-3}$, then $\mathcal{P}^{RC}_{\bar{n}}(O)>\mathcal{P}^{DM}_{\bar{n}}(O)$ when $\bar{n}\ge 0.99$.  Both of these examples show that increasing the mean photon number of the signal, $\bar{n}$, will eventually lead to a point where the mimic protocol performs better than the direct measurement protocol with a single idler detector.  Intuitively, this comes from the fact that as $\bar{n}$ increases, we have a greater probability to pick large amplitude coherent states in the mimic protocol.

\begin{figure}
\center{\includegraphics[width=8cm]
{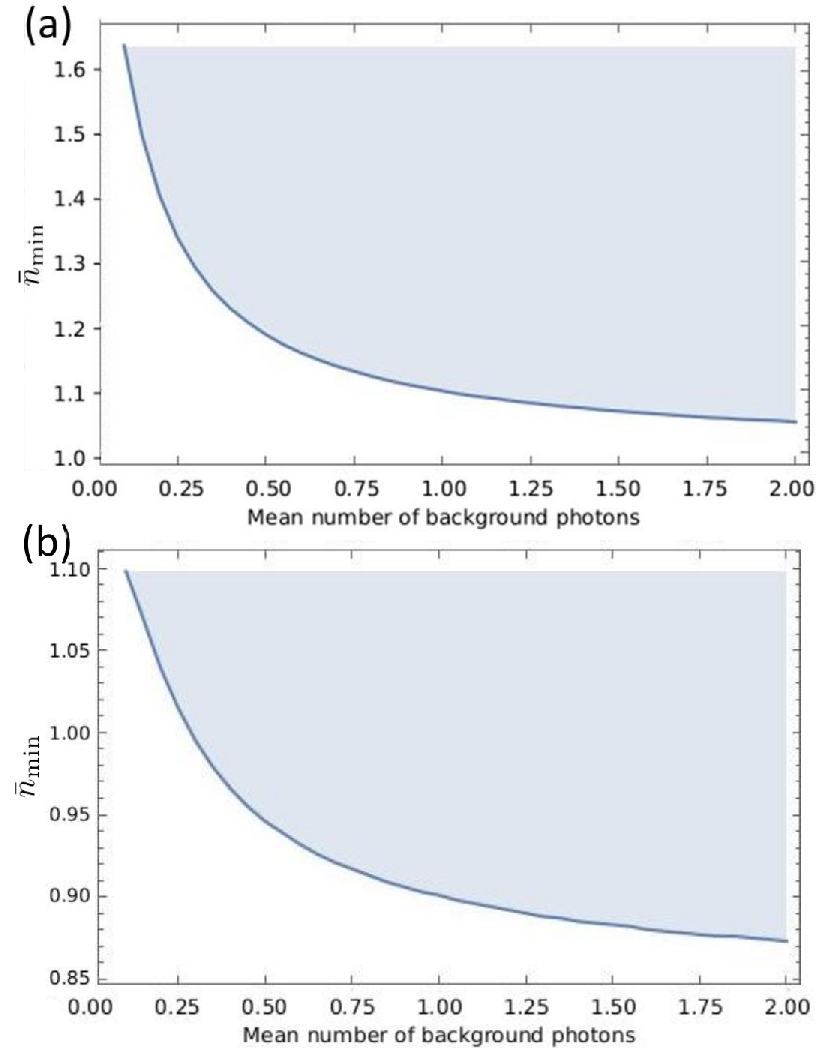}}
\caption{Parameter regimes for where mimic and 1 idler detector, direct measurement protocol each perform best as a function of $\bar{n}_B$. The quantity $\bar{n}_{\text{min}}$ plotted is the minimum value of $\bar{n}$ such that $\mathcal{P}^{RC}(O)>\mathcal{P}^{DM}(O)$.  In the shaded blue region the mimic protocol performs, on average, better than the 1 idler direct measurement protocol, for equal priors.  Both figures (a) and (b) are for $\kappa=0.1$ and the for the case where an object is present.  Figure (a) corresponds to $\eta=0.9$, while (b) is for $\eta=0.5$.}
\label{fig:plot3}
\end{figure}

The relations (\ref{rc_average}) and (\ref{ent_average}) can also be used to investigate how changing $\bar{n}_B$ affects the relative performance of the two protocols. In figure \ref{fig:plot3} we plot the minimum value of $\bar{n}$ for which $\mathcal{P}^{RC}_{\bar{n}}(O)>\mathcal{P}^{DM}_{\bar{n}}(O)$, for two different values of $\eta$. For brevity, we call this quantity $\bar{n}_{\text{min}}$.  In both figures, the shaded blue area denotes the region where $\mathcal{P}^{RC}_{\bar{n}}(O)$ is greater than $\mathcal{P}^{DM}_{\bar{n}}(O)$.  Figure \ref{fig:plot3} (a) is for $\eta=0.9$ and $\kappa=0.1$, while (b) is for $\eta=0.5$ and $\kappa=0.1$.  We see again that increasing $\bar{n}$ does eventually lead to the mimic protocol providing a better averaged probability.  
Both plots show that as $\bar{n}_B$ increases, $\bar{n}_{\text{min}}$ decreases. Increasing the background thus means that we achieve a greater relative performance of the mimic protocol for lower values of $\bar{n}$.  Comparing (a) and (b) shows that the detector efficiency has an effect on $\bar{n}_{\text{min}}$, which decreases for the smaller values for $\eta$.  A possible reason for this is that the direct measurement protocol requires detectors for both the idler and signal modes.  Inefficiencies in the idler mode decrease the probability for this detector to fire and also decrease the probability to enhance the signal mode's mean photon number.  This compounds the effects of the inefficiency of the signal detector.

\section{Conclusions}
\label{sec:conc}
We have presented an object detection protocol that uses random coherent states to mimic the operation of a direct measurement quantum illumination protocol.  In the so called mimic protocol, one transmits coherent pulses with a randomly chosen amplitude, $\alpha$. The mean photon number of each pulse will differ, but the averaged value is fixed.  Furthermore, the average ensemble of coherent states is a thermal state.  This ensures that the photon statistics of the signal mode is similar to the background.  This mimics the behaviour of a two mode squeezed vacuum, where the reduced state of the signal mode is also a thermal state.  Additionally, the random variations in the signal intensity mimics the measurement based conditioning for quantum illumination schemes that involve measures on the idler mode. 
The preparation of the random ensemble of coherent states could be achieved using electro-optical modulators in an approach that is similar to that which has been experimentally demonstrated already, for the task of covertly distributing information \cite{ghostdisplace}.

We compared the performance of the mimic protocol to coherent states with fixed amplitudes and a direct measurement scheme where one measures the idler mode to condition the signal mode.  To make the comparison fair, we use the same averaged mean photon number for each protocol.  It was found that the mimic protocol performed better than coherent states with fixed intensity.  For reflectance of 0.1 and a background mean photon number of 3.0, we found that as the signal mode mean photon number went to one, the performance of the mimic scheme became similar to the direct measurement scheme.  When the mean photon numbers of the signal mode became greater, the mimic scheme eventually performed better than the direct measurement scheme.  Numerical investigation for different parameters confirms that increasing the mean photon number will eventually result in the mimic protocol out performing the considered direct measurement protocols.  A particularly interesting example was for a reflectance of $10^{-7}$, where we found that both the mimic and direct measurement schemes could identify an object using of order $10^7$ pulses.

The relative performance of the direct measurement protocol with a single idler detector and the mimic protocol was investigated further.  We derived an analytic condition for comparing the two protocols. This applied in the case where one uses equal priors probabilities for the object being present or not.  The condition allows one to explore the parameters space to determine when one should use each protocol. For instance, we found that the direct measurement protocol will outperform the mimic protocol up to a particular value of the mean photon number, $\bar{n}_{\text{min}}$.  Furthermore, we found that increasing the number of background photons decreases the value of $\bar{n}_{\text{min}}$.

The protocol and presented results are also valid if we replace coherent pulses with phase-randomized coherent pulses.  We can again choose the intensity of these pulses randomly such that the average state is thermal. The results are then identical to the case of random coherent states as the phase of the coherent states is not relevant to the detection process.

The current paper establishes that some of the advantage from quantum illumination can be achieved without the need for twin-beams with correlations in the photon number. Instead, one can use coherent pulses with random intensities.  The results are of interest in situations where such higher intensities are needed.  This is for two reasons: (1) the mimic protocol has been shown to work best in this regime and (2) generating high intensity coherent states is more straightforward than generating either higher intensity TMSV states or twin-beam states.  This will be important in situations where we have very low reflectances and high background counts. 

\section*{Acknowledgements}
We acknowledge financial support from the United Kingdom (UK) Engineering and Physical Sciences Research Council for funding via the UK National Quantum Technology Programme and the QuantIC Imaging Hub (Grant No. EP/T00097X/1). We also acknowledge helpful discussions with D. K. L. Oi and Hao Yang.

\section*{Appendices}

\subsection*{A: Derivation of detection probability for the mimic scheme}
The probability for the detector to not fires, given an object is present, is given in Eq. (\ref{eq:detect}).  In this appendix we will explain how this probability is calculated.  Let $\hat U_{\kappa}$ be a unitary that describes the action of a beam-splitter with reflectance $\kappa$, as shown in figure (\ref{fig:model}).  This models the reflection of a signal state from the object and the injection of thermal photons.  The input states for the beam-splitter are a coherent state $|\alpha\rangle$ and the thermal state $\hat\sigma_{\bar{m}}$, where $\bar{m}=\bar{n}_B/(1-\kappa)$.  The output state is $\hat U_{\kappa}[|\alpha\rangle\langle\alpha|\otimes\hat\sigma_{\bar{m}}]\hat U^{\dagger}_{\kappa}$ To evaluate this we use the coherent state representation of a thermal state, given in (\ref{thermal2}).  Using linearity together with the well known properties of coherent states at a beam-splitter \cite{Loudon,mmqo}, we find that 
\begin{widetext}
\begin{eqnarray}
&&\hat U_{\kappa}\left[\,|\alpha\rangle\langle \alpha|\otimes\hat\sigma_{\bar{m}}\right]\hat U^{\dagger}_{\kappa}=\\
&&\frac{1}{\pi\bar{m}}\int e^{-|\beta|^2/\bar{m}}|\sqrt{1-\kappa}\alpha+i\sqrt{\kappa}\beta\rangle\langle\sqrt{1-\kappa}\alpha+i\sqrt{\kappa}\beta |\otimes|i\sqrt{\kappa}\alpha+\sqrt{1-\kappa}\beta\rangle\langle i\sqrt{\kappa}\alpha+\sqrt{1-\kappa}\beta|d^2\beta\nonumber.
\end{eqnarray}
\end{widetext}
The second mode is the one which is incident on the detector, while the first mode is the loss mode, which we trace out.  The detector mode is measured by an inefficient detector.  This is equivalent to an ideal threshold detector with a lossy channel in front of it. The effect of passing a coherent state through this lossy channel is: $|\alpha'\rangle\langle\alpha'|\rightarrow|\eta\alpha'\rangle\langle\eta\alpha'|$.  The probability to not detect a photon is 
\begin{widetext}
\begin{eqnarray}
P(0|O,|\alpha|^2)&=&\int \frac{\exp\left(-|\beta|^2/\bar{m}\right)}{\pi\bar{m}}\exp\left(-\eta|i\sqrt{\kappa}\alpha+\sqrt{1-\kappa}\beta|^2\right)d^2\beta.\nonumber\\
&=&\frac{\exp(-\eta\kappa|\alpha|^2)}{\bar{m}\pi}\int\exp\left(-A|\beta|^2+B\alpha\beta^*+B\beta\alpha^*\right)d^2\beta,
\end{eqnarray}
\end{widetext}
where 
\begin{eqnarray}
A=\frac{1+\eta\bar{m}(1-\kappa)}{\bar{m}},\nonumber\\
B=\eta\sqrt{\kappa(1-\kappa)}.
\end{eqnarray}
The coefficients $A$ and $B$ are both non-negative.  The above integral can be solved by separating $\beta$ into real and imaginary parts and performing the resulting Gaussian integrals.  After some algebraic manipulations, we find that
\begin{equation}
\label{appAfinal}
P(0|O,|\alpha|^2)=\frac{\exp(-\eta\kappa|\alpha|^2)}{1+\eta\bar{m}(1-\kappa)}\exp\left(\frac{\eta^2\kappa(1-\kappa)\bar{m}|\alpha|^2}{1+\eta\bar{m}(1-\kappa)}\right).
\end{equation}
The mean number of photons injected into the detector mode is $\bar{n}_B=\bar{m}(1-\kappa)$.  Using this relation in (\ref{appAfinal}) followed by some straightforward algebra will give Eq. (\ref{eq:detect}).

\subsection*{B: Method for including detector dark counts in $\bar{n}_B$}
\begin{figure}
\center{\includegraphics[height=5cm]
{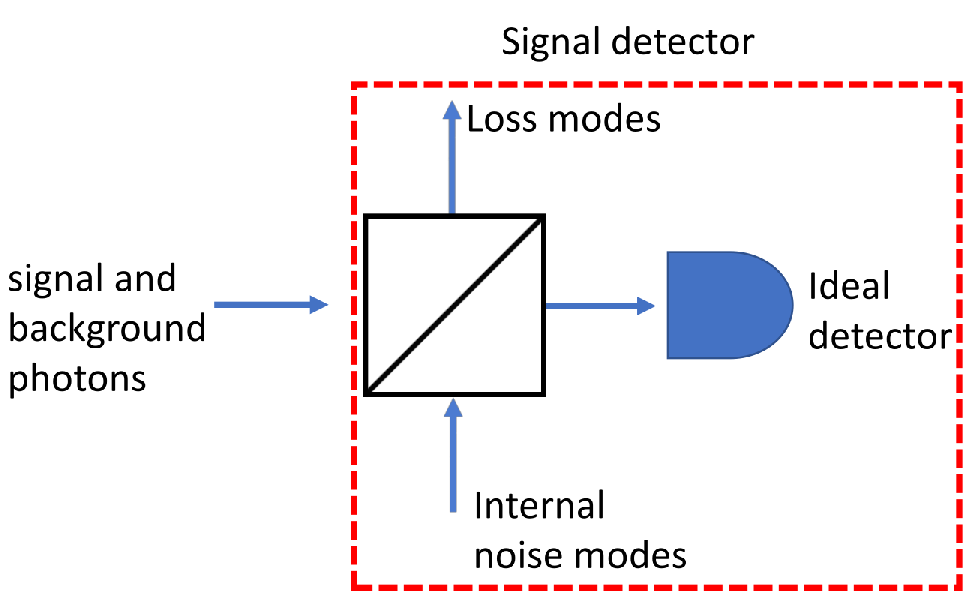}}
\caption{A figure showing the model for an inefficient detector with dark counts. We describe this as an ideal detector with a beam-splitter in front of it. The transmittance of the beam-splitter is $\eta$, the efficiency of the detector. Undetected photons are scattered into the loss mode, while dark counts result from photons from the internal noise mode.}
\label{fig:detector}
\end{figure}

In this appendix we outline how $\bar{n}_B$ can be modified to include the effects of dark counts.  First suppose that one blocks all the signal and stray background photons from entering the detector.  The detector still has a probability, $P_D$, to fire in each time-bin due to detector dark counts.  This can be modeled by changing our description of an inefficient detector.  In figure \ref{fig:detector} we outline a model of an inefficient detector with dark counts.  In the model, the detector is an ideal detector, but with a beam-splitter in front of it. The detector efficiency, $\eta$, corresponds to the probability for an incoming photon to be transmitted by the beam-splitter.  The effects of dark counts are described by an internal noise mode, which is in a thermal state with mean photon number $\bar{m}$. Suppose there is no incoming light (either signal or background); the only photons incident on the ideal detector are from the internal noise mode.  On average, in each time-slot, we have $\bar{n}_D=\bar{m}(1-\eta)$ mean photons incident on the detector from the internal noise mode.  

The dark count probability, $P_D$, can be found by first calculating the probability for the detector to not fire due to these internal noise photons.  This is found by using the model shown in figure \ref{fig:detector}, where the signal mode is the vacuum and the noise mode is a thermal state with mean photon number $\bar{m}=\bar{n}_D/(1-\eta)$.  A straightforward calculation shows that the probability to not register a click is $P_0=1/(1+\bar{n}_D)$, which means that
\begin{equation}
\label{pec}
P_{D}=\frac{\bar{n}_{D}}{1+\bar{n}_D}.
\end{equation}
This relates the probability of a dark count to the mean number of noise photons incident on the hypothetical ideal detector.  This is equivalent to injecting $\bar{n}_D/\eta$ background photons into the mode that is incident on the real (non-ideal) detector.  If we already had $\bar{n}'_B$ background photons in this mode, then the effects of detector dark counts are included by modifying the mean photon number of the background to $\bar{n}_B=\bar{n}'_B+\frac{\bar{n}_D}{\eta}$.  In practice, we won't know $\bar{n}_D$, but instead will know the dark count probability.  We would then use Eq. (\ref{pec}), which gives: $\bar{n}_D=P_D/(1-P_D)$.  The effective mean number of background photons incident on the real detector can be taken as
\begin{equation}
\label{modified_background}
\bar{n}_B=\bar{n}'_B+\frac{P_D}{\eta(1-P_D)}.
\end{equation}
This is the desired result.
  
\subsection*{C: Derivation of $\mathcal{P}^{RC}(O)$}
In this appendix we outline a derivation of (\ref{rc_average}), the averaged probability that an object is present given the results for a single measurement and that there is an object. The formal definition of $\mathcal{P}^{RC}(O)$ is given in Eq. (\ref{prc_def}).  From this Eq., we see that $\mathcal{P}^{RC}(O)$ is the sum of two integrals, which we denote as $I_1$ and $I_0$, where the subscript refers to whether we register a detection ($I_1$) or have no detection ($I_0$).  Using the detection probabilities (\ref{eq:detect}) and Bayes' rule, (\ref{bayes}), we find that
\begin{eqnarray}
I_0=\frac{1}{\bar{n}(1+\eta\bar{n}_B)}\int^{\infty}_0 e^{-\lambda/\bar{n}} \frac{\exp(-\gamma\lambda)^2 }{\exp(-\gamma\lambda)+1}d\lambda,\nonumber\\
I_1=\frac{1}{\bar{n}}\int^{\infty}_0 e^{-\lambda/\bar{n}} \frac{(1-\exp(-\gamma\lambda)/Z)^2 }{[1-\exp(-\gamma\lambda)/Z]+[1-1/Z]}d\lambda,
\end{eqnarray}
where $\lambda$ is the mean photon number of a pulse, $\gamma=(\eta\kappa)/(1+\eta\bar{n}_B)$ and $Z=1+\eta\bar{n}_B$.  Both of these integrals can be transformed to standard forms by suitable changes of variables.  For both integrals, we start by using the substitution $t=\exp(-\gamma\lambda)$.  For $I_1$, we can re-arrange to obtain
\begin{equation}
\label{i1_working}
I_1=\frac{1}{\bar{n}\gamma} AZ\int^{1}_0 \frac{t^{\beta-1}(1-t/Z)^2}{1-At}dt,
\end{equation}
where $\beta=1/(\bar{n}\gamma)$ and $A=(2Z-1)^{-1}$.  The integral representation of the hypergeometric function is \cite{AA}
\begin{eqnarray}
\label{h21}
&&\,_2 F_1(a,b;c;x)=\\
&&=\frac{\Gamma(c)}{\Gamma(b) \Gamma(c-b)}\int^{1}_0 t^{b-1} (1-t)^{c-b-1} (1-xt)^{-a} dt\nonumber,
\end{eqnarray}
where $c>b>0$ and  $\Gamma(y)$ are gamma functions. Using this, we can re-express (\ref{i1_working}) as a linear combination of hypergeometric functions
\begin{eqnarray}
\label{i1}
&&I_1 =\frac{1}{\bar{n\gamma}}\left(\frac{Z}{2Z-1}\right)\Big[\frac{\,_2 F_1(1,\beta;\beta+1;A)}{\beta}\\
&& -2\frac{\,_2 F_1(1,\beta+1;\beta+2;A)}{Z(\beta+1)}
+\frac{\,_2 F_1(1,\beta+2;\beta+3;A)}{Z^2 (\beta+2)} \Big]\nonumber.
\end{eqnarray}
The other integral, $I_0$, can be written as 
\begin{equation}
I_0=\frac{1}{\eta \kappa\bar{n}}\int^{1}_0 \frac{t^{\beta+1}}{t+1}dt.
\end{equation}
To simplify this further, we first multiply the denominator and numerator of the integrand by $(1-t)$ and then make a change of variable to $u=t^2$.  This gives the integral
\begin{eqnarray}
I_0&=&\frac{1}{2\eta \kappa\bar{n}}\int^{1}_0 \frac{u^{\beta/2}(1-u^{1/2})}{1-u}du\nonumber\\
&&=\frac{1}{2\eta \kappa\bar{n}}\int^{1}_0 \left(\frac{1-u^{(\beta+1)/2}}{1-u} -\frac{1-u^{\beta/2}}{1-u}\right)du.\nonumber\\
\end{eqnarray}
The Harmonic function, $H(x)$, is equal to $H(x)=\gamma_{euler}+\frac{d}{dx}\ln(\Gamma(x))$, where $\gamma_{euler}$ is Euler's number and $\Gamma(x)$ is the standard Gamma function \cite{AA}.  The Harmonic function also has a useful integral representation \cite{AA}
\begin{equation}
\label{hintegral}
H(x)=\int^{1}_0 \frac{1-u^x}{1-u}du.
\end{equation}
Using this we can re-express $I_0$ as
\begin{equation}
\label{i0}
I_0=\frac{1}{2\eta \kappa\bar{n}}\left[H\left(\frac{1}{2}+\frac{\beta}{2}\right)-H\left(\frac{\beta}{2}\right)\right].
\end{equation}
Some simple algebra shows that the sum of $I_0$ and $I_1$ is equal to (\ref{rc_average}).

\subsection*{D: Photon statistics of sub-ensembles}
In this appendix we discuss the photon statistics of the various states of the signal mode.  Recall, in both the mimic and direct detection schemes, the averaged state of the signal mode is a thermal state.  The averaged state is, however, realised in different ways for each protocol.  In the mimic protocol we prepare coherent states of different amplitudes with probabilities such that the average state is (\ref{thermal2}), while in direct measurement schemes, the signal state is prepared by measuring the idler mode of the state (\ref{tmsv}).  For simplicity, we limit our discussion to direct measurements schemes with a single idler detector.  This means that the signal mode is conditionally prepared in one of two states: $\hat\rho_0$ and $\hat\rho_1$, where the subscript denotes whether the idler detector has fired or not.  The form of the states $\hat\rho_i$ can be found using Eqs. (10) and (12) of \cite{yang2022}. The probability for the idler detector to not fire is $P_I(0)=(1+\eta\bar{n})^{-1}$, while the probability to register a click is $P_I(1)=1-P_I(0)$.  The averaged signal state is $P_I (0)\hat\rho_0 +P_I (1)\hat\rho_1 =\hat\sigma_{\bar{n}}$, which is a thermal state with mean photon number $\bar{n}$.  The mean photon number for $\hat\rho_0$ is $\bar{n}(1-\eta)/(1+\eta\bar{n})$, while the mean photon number for $\hat\rho_1$ is $\bar{n}+(1+\bar{n})/(1+\eta\bar{n})$.  The average of these is again $\bar{n}$, the mean photon number of the signal mode when we don't condition on an idler measurement.  Detecting light in the idler mode thus gives an enhancement in the signal mode mean photon number. In contrast, a failure to detect light in the idler mode suppresses the mean photon number in the signal mode.  However, the average behaviour is unchanged from the reduced state of the signal mode, as required by the no-signaling theorem \cite{nosignal}.  

\begin{figure}
\center{\includegraphics[height=10cm]
{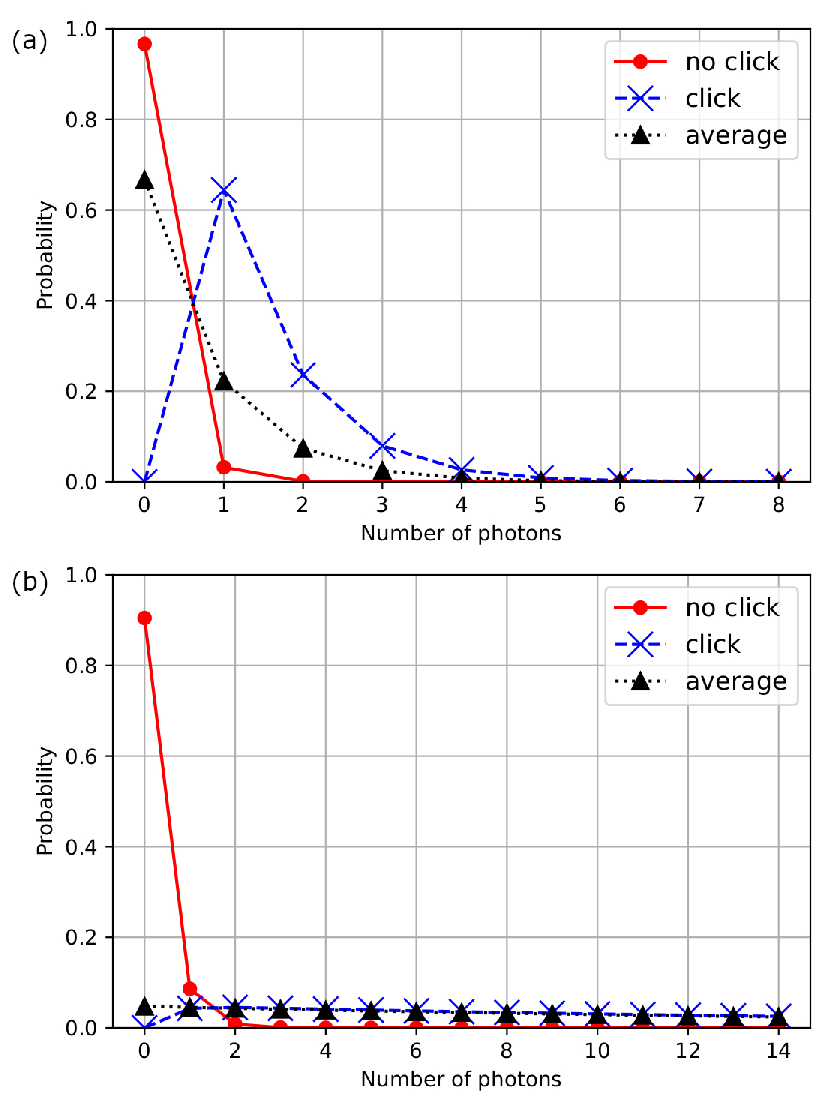}}
\caption{A plot showing the photon probability distribution for the conditional and average states of the signal mode in the direct measurement protocol. Both figures are for $\eta=0.9$, while (a) is for $\bar{n}=0.5$ and (b) is for $\bar{n}=20$.  In both figures, the triangles are for the average state, the circular dots are for the state conditioned on no click in the idler mode, while the x's are for the state conditioned on a click in the idler mode.}
\label{fig:appd1}
\end{figure}

We can gain more insight into the photon statistics by looking at some examples of the conditional photon statistics. 
In figure \ref{fig:appd1} we plot the conditional photon statistics for the direct detection scheme for $\eta=0.9$ and two different values of $\bar{n}$: $\bar{n}=0.5$ in (a) and $\bar{n}=20$ in (b).  In both (a) and (b) the triangles represent the photon statistics for the averaged state, the solid dots are for the conditional state $\hat\rho_0$ and the x's are for the conditional state $\hat\rho_1$.  For $\bar{n}=0.5$ the two conditional states have very distinct probability distributions, each of which differs from the average state. Heralding has a significant conditioning effect on the state when the detector fires. However, figure \ref{fig:appd1} (b) shows that for $\bar{n}=20$, the non-vacuum outcome for the probability distribution for $\hat\rho_1$ is almost identical to the averaged state. 
As the mean photon number is high, the click outcome is much more likely than the non-click (which is rarer but has a photon number reducing effect). Increasing $\bar{n}$ thus provides a small conditioning effect and keeps the photon probability distribution for $\hat\rho_1$ similar to that of the original thermal state. 

\begin{figure}
\center{\includegraphics[height=10cm]
{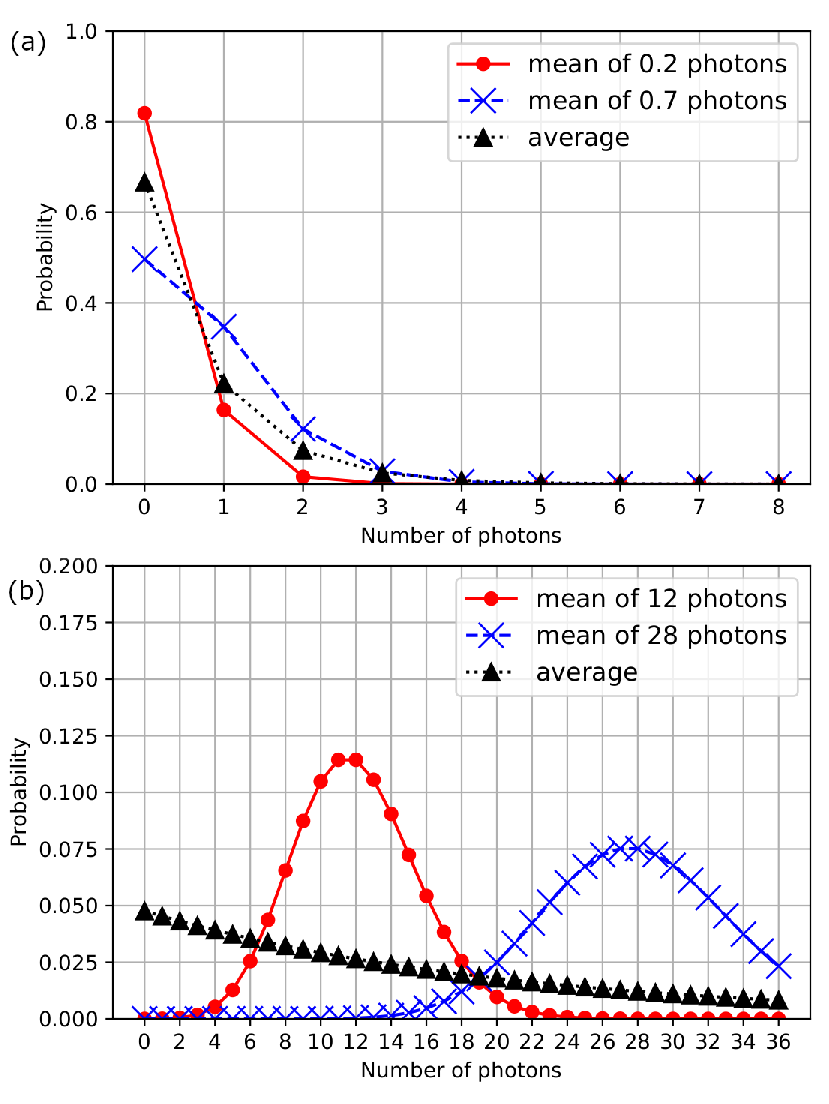}}
\caption{A plot showing typical photon probability distribution for coherent states and the averaged state of the signal mode. In figure (a) $\bar{n}=0.5$, the circular dots are for $|\alpha|^2=0.2$ and the x's are for $|\alpha|^2=0.7$.  In figure (b) $\bar{n}=20$, the circular dots are for $|\alpha|^2=12$ and the x's are for $|\alpha|^2=28$.  In both figures, the triangles denote the averaged state, which is a thermal state.}
\label{fig:appd2}
\end{figure}

In contrast, in the mimic protocol one transmits coherent states with Poissonian photon statistics.  With high probability, we will prepare coherent states with mean photon numbers, $|\alpha|^2$, that are close to the ensemble average $\bar{n}$.  Some examples of this are shown in figure \ref{fig:appd2}, where (a) is for $\bar{n}=0.5$ and (b) is for $\bar{n}=20$.  In both (a) and (b), the triangles denote the photon probability distribution for the averaged state, the dots are for probability distributions with a mean below $\bar{n}$ and the x's are for probability distributions with a mean above $\bar{n}$.  We see from (a) that for low values of $\bar{n}$, the resulting photon probability distributions are similar.  However, as $\bar{n}$ increases, we see from (b) that the distributions become more distinct. This feature helps explain why the mimic protocol performs better, in relative terms, as the mean photon number increases.  This analysis thus provides an intuitive understanding of the results found in section \ref{sec:compare}.

\end{document}